# Towards Photonics Enabled Quantum Metrology of Temperature, Pressure and Vacuum

Zeeshan Ahmed[1], Nikolai N. Klimov[1,2], Kevin Douglass[1], Jim Fedchak[1], Julia Scherschligt[1], Jay Hendricks[1], Jacob Ricker[1], and Gregory Strouse[1]

[1]Physical Measurement Laboratory, National Institute of Standards and Technology, Gaithersburg, MD, USA

[2]Joint Quantum Institute, University of Maryland, College Park, MD, USA

## 1. Introduction

SI realizations for pressure, temperature and vacuum are traditionally based upon legacy artifacts that exploit unique properties of materials. These material properties were traditionally probed using voltage/resistance measurements to enable sensitive measurement capabilities for well over a century. For temperature metrology, the platinum resistance thermometer was first proposed by Sir Siemens in 1871 with Callender developing the first device in 1890 [1]. Since that time incremental progress in the design and manufacturing of thermometers has delivered a wide range of temperature measurement solutions. The standard platinum resistance thermometer (SPRT) is the interpolating instrument for realizing ITS-90 temperature scale and its dissemination using resistance thermometry. Today, resistance thermometry plays a crucial role in various aspects of industry and everyday technology ranging from medicine and manufacturing process control, to environmental process control and the oil-and-gas industry [2-5]. Although industrial resistance thermometers can measure temperature with uncertainties as small as 10 mK, they are sensitive to mechanical shock, thermal stress and environmental variables such as humidity and chemical contaminants. Consequently resistance thermometers require periodic (and costly) off-line recalibrations [3].

Pressure metrology has followed a similar development curve starting with Torricelli's pioneering experiments in 1643. Improvements in mercury manometers since Torricelli's day have been made by incrementally improving column height measurement accuracies. Today, mercury manometers are still used at eleven National Metrology Institutes, including NIST. The NIST 13 kPa, 160 kPa and 360 kPa Ultrasonic Interferometer Manometers (UIMs) use ultrasound pulses to determine column heights with a resolution of 10 nm or a pressure resolution of 3.6 mPa and provide the world's lowest pressure uncertainties [4, 6, 7]. The UIM however, relies on elemental mercury, a dangerous neurotoxin slowly being phased out of modern economy. The desire to eliminate mercury along with increasing need for portable pressure standards has prompted considerable research in the search for alternative technologies.

The challenges faced by legacy metrology are foundational and as such, the field of metrology is ripe for disruption by new technologies that fundamentally re-imagine the realization, dissemination and measurement of SI units. A potentially disruptive technology for temperature, pressure, and vacuum metrology is photonics, the use of light-matter interaction to accurately measure physical properties. For example, light-matter interaction in optical cavities can be used to measure changes in a material's refractive index caused by changes in temperature or pressure. These innovative sensors have the potential to leverage advances in frequency metrology and telecom light sources to provide cost effective measurement solutions. A particularly exciting development is the possibility of using quantum sensors and/or quantum metrology tools such as the use of entangled photons or squeezed light to achieve sub-shot noise detection limit. Such a development would enable Heisenberg-limited metrology over a network scale using sensors that function as standards too.

Here we present the results of our efforts in developing novel photonic sensors for temperature, pressure and vacuum measurements. Our preliminary results indicate that using photonic devices such as ring resonators, photonic crystal cavities and Fabry-Pérot cavities, we can achieve measurement capabilities that are on par or better than the state-of-the-art in resistance metrology. The use of novel quantum phenomenon, such as Bose-Einstein condensates in vacuum, science holds the possibility of delivering an *ab initio* extreme high vacuum (XHV) standard that uses loss rate measurement of an atom trap to deliver unprecedented measurement of vacuum.

## 2. Thermometry

In recent years a wide variety of novel photonic thermometers have been proposed including photosensitive dyes [8, 9], hydrogels [10-12], fiber Bragg grating (FBG) [13-15], and on-chip integrated silicon photonic nanostructures [2, 16-18] running . At NIST, our efforts are aimed at developing novel photonic-based sensors and standards capable of outperforming resistance-based standards. Our goal is to develop a low-cost, readily-deployable, novel temperature sensor that is easily manufactured using existing technologies, such as CMOS-compatible manufacturing. The gamut of technologies under consideration runs from macroscale sapphire-based microwave whispering gallery mode resonator (WGMR) to micron-scale fiber Bragg gratings, to novel nanoscale silicon photonic crystal cavities.

**Sapphire WGMR:** Our earliest work focused on the development of sapphire WGMR devices. For the last 40 years, WGMRs fabricated with a wide variety of materials, including polysterene, fused quartz, glass spheres and mono-crystalline synthetic sapphire have been demonstrated as ultra-stable resonators [19, 20]. Considerable effort has been expended to eliminate thermal drift of the WGMR resonance frequency to improve their frequency stability [21, 22]. We exploited the large temperature dependence of sapphire WGMR's frequency and the ease of measuring peak frequency of high-$Q$ (quality-factor) modes to develop highly sensitive and accurate temperature sensors. Using a disk of monocrystalline sapphire we demonstrated a measurement uncertainty ($k$ = 2) of 0.01 °C, over the range of 0.01 °C to 100 °C [20]. Sapphire WGMR thermometers with spherical and cylindrical geometries were also been demonstrated. Our recent work is focused on developing compact, cylindrical devices that are mechanically stable and retain high Q's [23].

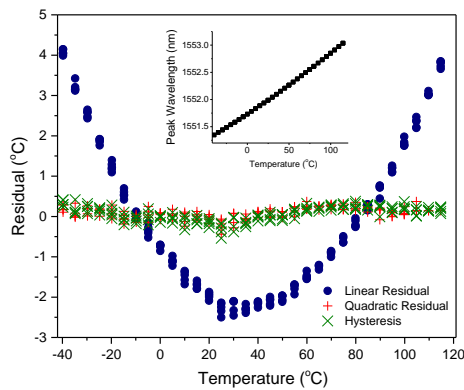

**Figure 1:** FBG show a quadratic dependence on temperature over the range of -40 °C to 120 °C.

**FBG Thermometry**: FBGs are narrow band filters commonly used in the telecommunications industry for routing information. FBGs are fabricated using photo-sensitive optical fibers (e.g., H$_2$-loaded Ge-doped fibers) that are exposed to spatially varying UV light. This modifies the local structure of silica, creating a periodic variation in the local refractive index, which behaves like a Bragg grating [13, 24]. The wavelength of light that is resonant with the Bragg period is reflected back, while non-resonant wavelengths pass through the grating. The grating equation is given by:

$$\lambda_B = 2n_e L$$

where $n_e$ is the effective refractive index, $L$ is the grating period and $\lambda_B$ is the Bragg wavelength. Changes in surrounding temperature impacts the effective grating period due to linear thermal expansion and the thermo-optic effect (refractive index due to temperature), resulting in a wavelength shift of ≈ 10 pm/°C [13, 24]. In a recent study, we carefully determined the impact of various sources of uncertainty on FBG-based measurements. Our results demonstrate that for an unstrained, "aged" sensor in a humidity-controlled environment, over the temperature range of -40 °C to 120 °C, the FBG sensor shows a quadratic dependence on temperature (Fig. 1) with a measurement uncertainty of ≈ 0.5 °C, comparable to Type J thermocouples [25]. Embedding FBG, into structures such as 3D printed scaffolds [26] or aerospace-grade composite materials [27, 28] can provide useful information on the structural properites of the material. However, as we have shown recently cross-talk between temperature and strain response of FBG sensors is significant and limits their usefulness in such applicaitons [26].

**Silicon Photonic Thermometers:** Silicon photonic thermometers are another class of photonic thermometers that exploit a material's thermo-optic coefficient (TOC) to achieve high temperature sensitivity. Silicon's TOC is ≈10x larger than silica. These micro- and nanoscale silicon sensors are fabricated out of very thin (≈ 220 nm) crystalline silicon (Si) layers using conventional silicon-on-insulator (SOI) CMOS-technology. Photonic devices and structures are shaped from the topmost crystalline Si layer of SOI wafer via photo- or electron beam lithography, followed by an inductive plasma reactive ion etch. At the end of the fabrication process, the devices are usually top-cladded with a thin poly(methyl methacrylate) (PMMA) or silicon dioxide ($SiO_2$) protective layer.

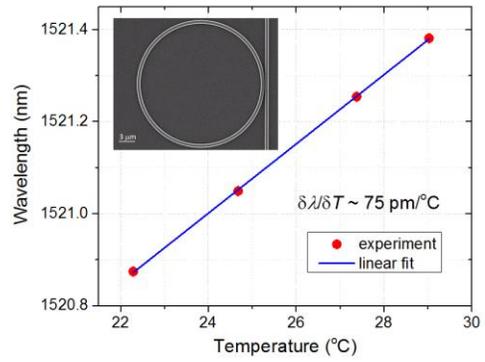

**Figure 2:** Temperature dependent response of silicon ring resonator shows a temperature sensitivity of ≈75 pm/K. Insert shows an SEM image of the ring resonator device

We have recently fabricated a wide variety of photonic temperature sensors, including Si waveguide Bragg reflectors (Si WBG)[29], ring resonators[2, 30-32], and photonic crystal cavities[33]. In a recent study, we examined the temperature dependence of Si WBG over the temperature range of 5 °C to 160 °C. We observed a temperature sensitivity of ≈80 pm/°C, a factor of eight improvement over FBG devices. Our results indicate a combined expanded measurement uncertainty ($k = 2$), 1.25 °C, is dominated by the uncertainty resulting from peak center measurement [29].

Lower uncertainties could be achieved in narrower band stop devices, such as ring resonators and photonic crystal cavities. We recently demonstrated that silicon-based optical ring resonators can be used for ultra-sensitive thermal measurements. In particular, we showed that a device with a ring diameter of 11 μm and gap (between waveguide and ring) of 130 nm can be used as a photonic thermometer (Fig. 2) with temperature resolution of 1 mK and noise floor of 80 μK at integration times of 1 second [2]. Our device optimization work with ring resonators indicates a zone of stability in the parameter space (*waveguide width* > 600 nm, *waveguide-ring gap* ≈130 nm and *ring radius* >10 μm), where devices are less suscptible to routine fabrication errors. Consequently, these devices consistently show Q ≈$10^4$ and Δλ/ΔT ≈80 pm/°C.[30, 32]

Similarly high temperature sensitivity is achieved using photonic crystal cavity (PhCC) devices [33]. A typical example of PhCC device, shown in Fig 3, consists of an 800 nm wide silicon waveguide patterned with a one-dimensional array of subwavelength holes (hole diameters range from 170 nm to 200 nm) that acts as a resonant cavity for ≈1550 nm light. As shown in Fig 3, Q ≈ 31,000 is readily achieved with a temperature dependence of $\delta\lambda/\delta T \approx 70$ pm/°C for PMMA-cladded devices and ≈80 pm/°C for $SiO_2$- cladded devices. Temperature resolution can be further improved in the future by fabricating higher Q devices.

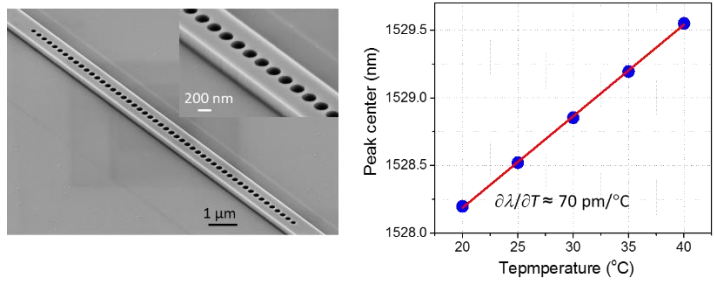

**Figure 3:** Fabry-Perot cavity-based silicon photonic thermometers (a) SEM image of uncladded Si PhC-C thermometer. (b) Temperature dependence of the resonance wavelength of PMMA-cladded Si PhC-C.

Our future work will focus on utilizing the PhCC design to fabricate opto-mechanical devices. In opto-mehcanical devices, the optical field of the cavity is coupled to its mechanical motion, providing a convienent means of probing the cavity dynamics.[34] The coupling of opto-mechanical modes is sensitive to photon occupation of the nanoresonator and can be used to directly measure the thermodynamic temperature of the mechanical mode directly. These devices would provide a convienent means of realizing thermodynamic temperature outside the lab and by eliminating the need for a traceability chain, opto-mechanical devices of this type could lead to significant reductions in the cost of sensor ownership.

## 3. Quantum-Based Primary Pressure Standard

The future of pressure and vacuum measurement will rely on lasers and Fabry-Pérot optical cavities measuring fundamental physics of light-matter interaction. The dielectric susceptibility of matter is greater than the dielectric susceptibility of vacuum. As a result, light travels more slowly and exhibits a shorter wavelength in gas than it does in vacuum. The non-resonant interaction of light and matter, i.e. the refractive index, forms the basis of a bold and innovative approach being undertaken at NIST to disruptively change the way we realize and disseminate the SI unit of pressure, the pascal.

The underlying metrology leverages advances in length metrology to make ultra-accurate determinations of the refractive index of a gas at a given pressure. Physical measurements are then combined with the calculated molar refractive index of the gas at standard conditions using *ab-initio* quantum chemistry calculations. This methodology yields a quantum-based primary pressure standard. The heart of the measurement apparatus is an innovative dual Fabry-Pérot cavity system that utilizes two Fabry-Pérot cavities built on a single low-expansion glass spacer, with one of the cavities serving as a vacuum reference [35]. Figure 4 shows the experimental set up for a dual Fabry–Pérot device and the working prototype.

Our preliminary measurements demonstrate a pressure resolution of 0.1 mPa ($7.5\times10^{-7}$ Torr), outperforming the NIST ultrasonic interferometer manometer (currently the most accurate manometer in the world) by *35X*. Additionally, the lowest pressure measured is 10X more sensitive (1 mPa vs. 10 mPa). Uncertainty of the photonic based pressure standard varies between 0.02% at medium vacuum (1 kPa), to 35 parts in $10^6$ at atmospheric pressure (100 kPa), with repeatability of 5 parts in $10^6$ or better. This indicates that the standard, once fully developed, will fundamentally change the way the SI unit for pressure is realized and disseminated, with the associated benefits of being mercury free, and delivering lower uncertainties than the existing primary pressure mercury manometers. A particularly exciting aspect is that the photonic-based pressure standard is smaller and more compact than existing mercury manometers and has the potential to become a standard reference instrument for realizing the pascal.

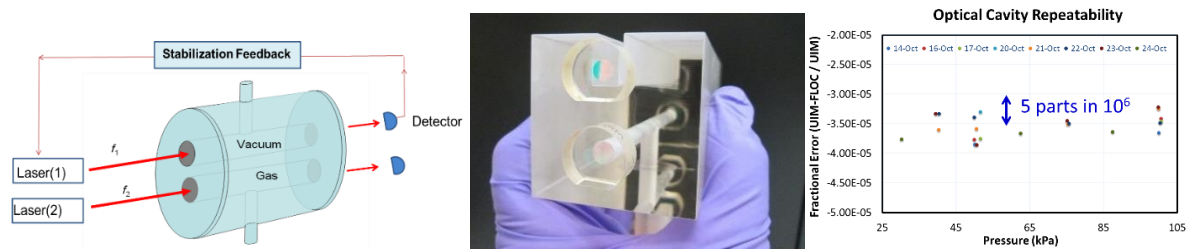

**Figure 4:** Above Left: A fixed length optical cavity (FLOC) realizes pressure by comparing the beat-frequency, $f_1$-$f_2$, as the pressure in gas cavity is increased while the vacuum in the reference cavity is held constant. This standard is simple (has no moving parts) and requires that the temperature of the cavity and molar index of the gas in the cavity is known from either measurements (for example nitrogen) or from theory (helium). Above Middle: Working prototype of a fixed length optical cavity (FLOC). A laser is locked to the top cavity which is filled with a gas, while a second laser is locked to the bottom cavity held at high vacuum. Above Right: Repeatability of the fixed length optical cavity is 5 parts in $10^6$ over the range of 30 kPa to 100 kPa. This is comparable to the mercury manometer standards.

**4. Dynamic Measurement of Pressure**: The dynamic measurement of pressure[a] is critical for a wide variety of industrial processes, transportation, human health, ballistics, and the environment. For example, gasoline and diesel combustion engines experience maximum combustion pressures of 0.1 MPa to 10 MPa [36, 37], and advanced engine concepts have been proposed for up to 50 MPa. Gas turbines, used in aircraft and for power generation, operate at up to 4.5 MPa with a frequency spectrum of several hertz to 30 kHz.[38] Accurate measurement of dynamic pressure for combustion supports improved optimization of fuel efficiency. Explosive detonation is used to launch projectiles in weapons (e.g. cannons, firearms, missiles), and the dynamic pressure is key in determining projectile range and service life of the gun barrel. A major cause of traumatic brain injury for soldiers comes from exposure to blast waves, which typically have peak pressures of 1 MPa and rise times on the order of microseconds [39-41]. Medical applications of dynamic pressure, such as blood pressure measurements, have a range of 10 kPa to 100 kPa, with a frequency content of the order of 100 Hz (a 1 % time slice of a heart beat) [42]. Other applications include processes such as rapid cooling of high voltage circuit breakers, deployment of automotive airbags, hydraulic fracking, hydraulic and pneumatic fluid pressure for instrument control or manufacturing (e.g., waterjet cutting), and injection molding [43].

---

[a] The terms dynamic pressure and static pressure used in this paper are the conventional ones used in the metrology community, see for example [30, 35-38]. In metrology, static and dynamic (for any quantity) refer to the time-varying nature of the quantity. A pressure is "static" if it remains constant for an amount of time on the order of the measurement time, whereas the pressure is "dynamic" if its value varies significantly during the measurement. It should be recognized that there exists another definition for static and dynamic pressure which is used in fluid mechanics, that expresses information about the velocity of the flow field. In fluid mechanics, the static pressure is the pressure in a flow field at rest if the observer is traveling at the speed of the moving fluid, and the dynamic pressure is the kinetic energy of the moving fluid. This latter definition is not used in this paper.

Unlike static or steady state pressure, there are no commercially-available SI-traceable calibration services for dynamic pressure, therefore the quantifiable uncertainties of dynamic pressure sensors and measurements and unknown [43]. There are currently no National Metrology Institutes that list a Calibration Measurement Capability (CMC) for dynamic pressure in the BIPM Key Comparison Database. An assumption is made that the sensor will respond the same to a dynamic pressure as it would to a steady-state pressure of the same magnitude, as long as the sensor's natural frequency far exceeds that of the generated pressure. The time-dependent response is assumed to be linear. Validating this assumption requires a dynamic pressure source of known (traceable) performance, which leads to a circularity of the argument; the only way to verify the capability of the source is through measurement with dynamic sensors, whose calibration depends on the performance of the source, as so forth.

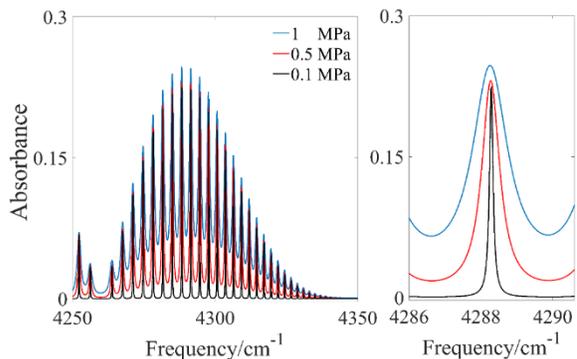

**Figure 5:** The plot shows simulated spectrum of the R branch of the 2←0 vibrational band of carbon monoxide at various pressures in a 5 cm path length and 10 % partial pressure. The panel on the right is a zoom on R(7).

We are developing a pressure standard for dynamic measurements using line-of-sight absorption spectroscopy to measure dynamic pressure and temperature in a gas. This relies on the same traceability principle first proposed in [44], using the unique quantum mechanical characteristics of the molecules as the standard for pressure. As an example, a portion of the v = 2←0 vibrational band of carbon monoxide is shown in Fig 5, and illustrates the pressure dependence of the absorption transitions at a temperature of (25 °C) and 10 % partial pressure.

Our method builds on the vast body of work in the gas sensing and combustion diagnostics fields [45-48]. Absorption spectroscopy using narrow linewidth cw lasers in a shock tube was first demonstrated over two decades ago [49-51]. The recent review by Hanson describes the current state-of-the-art laser diagnostic techniques for combustion and chemical kinetics research [45]. The primary method we are pursuing is based on measuring pressures in the gas phase. Depending on the application, the relevant pressures range from 10 kPa up to several hundred megapascals, frequencies range from one hertz to hundreds of kilohertz, and fluids are both liquid and gas. As pressure is increased, the measurement of the absorption lineshape profile becomes more challenging. Thus the technique is most likely to be successful up to a few megapascals. A successful primary optical standard can be compared over its range to calculated pressure sources and pressure sensors; validation of (or correction to) the source and sensor models can give more confidence in using them over extended ranges and underconditions for which the optical method is not presently anticipated to be used.

The path to a traceable dynamic pressure standard requires four steps: 1) Calibrate a reference static pressure sensor against a primary static pressure standard, such as a piston gauge or a mercury manometer. 2) Measure lineshapes of candidate gas absorbers (such as R(7) for CO) at a series of known pressures (using the calibrated reference sensor from step 1), known temperatures, and known gas concentrations under *static* conditions. Some of this data already exists for relevant pressure and temperatures in databases such as HITRAN. Identify appropriate lineshape model for data analysis. 3) Under dynamic conditions using the spectroscopic technique proposed herein, measure the lineshape and temperature. 4) Fit the measured data to line-shape models with pressure as the fitting parameter and temperature determined from the ratio of two line intensities. In essence, traceability in this method is to a static primary pressure standard, with the assumption that the molecules achieve thermal equilibrium faster than the time scales imposed by the dynamic pressure, such that the lineshapes are the same measured dynamically as they would be statically. The molecular lineshapes become the transfer standard between the static primary pressure standard and the dynamic pressure standard. The molecular absorption measurement route also lends itself towards allowing dissemination of the measurement scheme via the use of a standard reference database (SRD) such as HAITRAN.

## 5. Photonic-based Vacuum metrology

Ultra-high vacuum (UHV) is roughly defined as pressures below $10^{-6}$ Pa. Presently, most practical Earth-bound applications of vacuum require pressures no lower than about $10^{-10}$ Pa (XHV). In that pressure regime, the mean free path of molecules of atoms are longer than 1 km, the gas density is on the order of $10^8$ cm$^{-3}$ and lower, and we tend to speak of low pressure in terms of the reduction of gas density rather that in the reduction of mechanical force. Scientists and engineers with systems operating in the UHV and below are typically concerned about reducing contamination, increasing mean free paths, or the detection of ions or molecules. Indeed, many applications in advanced research and technology critically depend on vacuum environments in the range of $10^{-6}$ Pa to $10^{-10}$ Pa, including accelerator facilities, the space sciences, nanotechnology, semiconductor processing, surface science, gravitational wave detectors, etc. To date, the most common sensors for vacuum are ionization gauges and quadrupole mass analyzers, which are also based on ionization technology. These sensors fundamentally require calibration and can alter the very environment they are trying to measure by outgassing, pumping gas, and catalyzing chemical reactions.

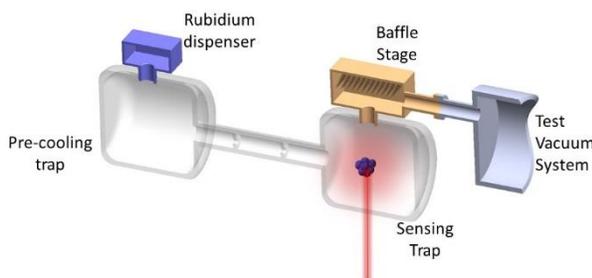

**Figure 6:** Schematic of the cold-atom vacuum standard (CAVS)

Photonic-based sensors are less intrusive to the vacuum environment and could serve as absolute primary sensors that would not require calibration. Multiphoton ionization techniques have been employed to measure the gas density for particular gas species, such as CO [52, 53]. However, what is greatly desired is an absolute sensor that can detect the most common background gases in the vacuum such as $H_2$, $N_2$, $CO_2$, etc., and for process gases like Ar or He. In the UHV, $H_2$ tends to dominate the background. However, traditional spectroscopy techniques are difficult to apply to $H_2$ (and most of the above mentioned gas species) because the resonance lines tend to lie in the ultraviolet, and the gas densities are simply too small. On the other hand, using ultra-cold atoms to sense the vacuum level holds great promise as an absolute sensor in the UHV. This technology is sensitive to all gas species, is an absolute primary sensor, and can be made small and portable. Moreover, presently there exist no absolute sensors of UHV.

The proposed photonic sensor is based on the simple idea that the loss rate of ultra-cold atoms in, for example, a shallow magnetic trap, is due to collisions of the trapped ultra-cold atoms with gas in the vacuum. Fig 6 represents a potential prototype cold-atom vacuum standard (CAVS). Sensor atoms, such as Rb, are laser-cooled to sub-millikelvin temperatures and trapped in the pre-cooling trap region. The atoms are subsequently transferred to a sensing trap region using a pusher-laser, where the sensor atoms are trapped in a magnetic or dipole trap. The vacuum level of the test vacuum system is sensed in the sensing trap region. The number of atoms in the trap are detected using photonic techniques, such as laser-induced florescence. The pressure of the test vacuum system is determined from the loss rate of Rb atoms from the sensing trap. A series of baffles prevents un-trapped room-temperature atoms from entering the sensing trap region, and another baffle stage serves to condense any remaining Rb atoms and prevent them from entering the test vacuum system. It is possible to miniaturize the CAVS technology to a robust chip-scale package, potentially leading to an SI-traceable absolute vacuum detector that would be capable of withstanding rugged environments, such as space-flight conditions or for direct use in a processing facility. NIST presently has a nascent program to develop an SI-traceable CAVS, and a program at the University of British Columbia is presently investigating using cold-atoms as a vacuum sensor [54]. Early on, it was recognized that the loss-rate of trapped cold-atoms depended on the vacuum pressure (see Ref [55], for example), and there have been preliminary investigations into using magneto-optical traps as vacuum sensors [56, 57] but to date, no one has developed a deployable SI-traceable sensor based on cold-atoms. In the future, we will build upon these advances in photonic metrology to extend the cold-atom sensing platform to other quantities of interest beyond vacuum, most notably gyroscopic and inertial sensing.

**6. Summary:** The foundational limitations of legacy-based metrology has spurred significant interest in developing photonics-based sensors and standards. In this paper we have presented an overview of ongoing photonics research at NIST in the area of temperature, pressure and vacuum. Our proposed measurement solutions run the gamut from silicon photonic thermometers, to high accuracy Fabry-Pérot cavity based photonic pressure standard, to magneto-optic atom-traps for vacuum standard. The photonics-based metrology technologies under-development now leverage decades of research and development in telecom light sources, laser spectroscopy, and atomic physics, to enable a disruptive change in how we realize, disseminate and measure thermodynamic quantities in the age of ubiquitous sensing.

**Acknowledgement**

The authors acknowledge the NIST/CNST NanoFab facility for providing opportunity to fabricate silicon photonic temperature sensors.